\documentclass[12pt]{amsart} 
\usepackage{euler}
\usepackage{euscript}
\usepackage{multicol} 
\usepackage{amssymb}
\usepackage{amsmath} 
\usepackage{times} 

\newcommand{\tr}{\mbox{\upshape tr\ }}
\newtheorem{theorem}{Theorem}[section] 
\newtheorem{lemma}[theorem]{Lemma}

\newtheorem{corollary}[theorem]{Corollary}
\newtheorem{definition}[theorem]{Definition}
\numberwithin{equation}{section}
  \newcommand{\rz}{{\mathbb{R}}} 
  \newcommand{\nz}{{\mathbb{N}}} 
\newcommand{\e}{{\mathrm{e}}}
\def\tr{\mathop{\mathrm{tr}} \nolimits} 
\def\trG{\mathop{\mathrm{tr}}}

\newcommand{\idG}{{\boldsymbol 1}_{\boldsymbol G}}
\newcommand{\bsG}{{\boldsymbol G}}  

\newcommand{\cK}{{\mathcal{K}}}
\newcommand{\cL}{{\mathcal{L}}}

\IfFileExists{amsthm.sty}
{ 
  \usepackage{amsthm}
  \def\qed{\hbox {\hskip 1pt \vrule width 6pt height 6pt depth 1.5pt
        \hskip 1pt}}
}
{ 
\def\qed{\hbox {\hskip 1pt \vrule width 6pt height 6pt depth 1.5pt
        \hskip 1pt}}
\newenvironment{proof}{\textit{Proof.}}{\qed \flushright}
}
\newcommand{\LTcl}{L^{\mbox{\footnotesize\upshape cl}}}
\begin{document}
\title[Lieb-Thirring constants]
{New  bounds on the Lieb-Thirring constants}
\author[D. Hundertmark, A. Laptev and T. Weidl]
{D. Hundertmark$^1$, A. Laptev$^2$ and T. Weidl$^{2,3}$}
\begin{abstract}
Improved estimates on the constants $L_{\gamma,d}$, 
for $1/2<\gamma<3/2$, $d\in N$ in the inequalities for 
the eigenvalue moments of Schr\"{o}dinger operators are established.
\end{abstract}

\email{
\sffamily 
hdirk@princeton.edu,
laptev@math.kth.se,
weidl@math.kth.se} 

\maketitle
 
\maketitle

\section{Introduction}

Let us consider a Schr\"odinger operator in $L^2({\Bbb R}^d)$
\begin{equation}\label{Schr}
-\Delta + V,
\end{equation}
where $V$ is a real-valued function. 
The inequalities
\begin{equation}\label{LTh}
\tr (-\Delta + V)_-^\gamma\leq L_{\gamma,d}
\int_{{\Bbb R}^d}  V_-^{\gamma+\frac{d}{2}}\,dx\,,
\end{equation}
are known as Lieb-Thirring bounds and 
hold true with finite constants $L_{\gamma,d}$
if and only if $\gamma\ge1/2$ for $d=1$, $\gamma>0$ for $d=2$
and $\gamma\ge0$ for $d\ge3$. 
Here and in the following, $A_{\pm}=(|A|\pm A)/2$ denote 
the positive and negative parts of a self-adjoint operator $A$.
The case $\gamma> (1-d/2)_+$ was shown
by Lieb and Thirring in \cite{LT}. The critical case 
$\gamma=0$, $d\ge3$ is known as the Cwikel-Lieb-Rozenblum 
inequality, see  \cite{C,L,R} and also 
\cite{LY,Con}. The remaining case $\gamma=1/2$, $d=1$ 
was verified in \cite{W1}. 

It is  known that as soon as $V\in L^{\gamma+d/2}({\Bbb R}^d)$
and  the constant $L_{\gamma,d}$ is finite, then we have  Weyl's 
asymptotic formula  
\begin{align}\notag
\lim_{\alpha\to+\infty}
\frac{1}
{\alpha^{\gamma+\frac{d}{2}}}\,\tr (-\Delta+\alpha V)^\gamma_- &= 
\lim_{\alpha\to+\infty}
\frac{1}{\alpha^{\gamma+\frac{d}{2}}}
\iint_{{\Bbb R}^d\times{\Bbb R}^d}
    (|\xi|^2+\alpha V)_-^\gamma\,\frac{dxd\xi}{(2\pi)^{d}}\\
\label{asyformu}
&=L^{\mbox{\footnotesize\upshape cl}}_{\gamma,d}
\int_{{\Bbb R}^d} V_-^{\gamma+\frac{d}{2}}\,dx\,,
\end{align}
where the so-called classical constant
$L^{\mbox{\footnotesize\upshape cl}}_{\gamma,d}$ 
is defined by
\begin{equation}\label{Lgd}
L^{\mbox{\footnotesize\upshape cl}}_{\gamma,d}= (2\pi)^{-d} \int_{{\Bbb
R}^d} (|\xi|^2-1)_-^\gamma\,d\xi =
\frac{\Gamma(\gamma+1)}
{2^d\pi^{d/2}\Gamma(\gamma+\frac{d}{2}+1)}\,,
\quad \gamma\geq 0\,.
\end{equation}
This immediately implies $L_{\gamma,d}^{cl} \le L_{\gamma,d}$.

Until recently the sharp values of $L_{\gamma,d}$ were known
only for $\gamma\ge3/2$, $d=1$, (see \cite{LT,AL}), where 
they coincide with $L_{\gamma,d}^{cl}$. 
In \cite{LaWe} Laptev and Weidl extended this result to 
all dimensions. They proved that 
$L_{\gamma,d} = L_{\gamma,d}^{cl}$, for $\gamma\ge3/2$, $d\in{\Bbb N}$.
Recently, Hundertmark, Lieb and Thomas showed 
in \cite{HLT} that 
the sharp value of $L_{1/2,1}$ is equal to  $1/2$.

The purpose of this paper is to give some new bounds on 
the constants $L_{\gamma,d}$ for $1/2<\gamma<3/2$ and 
all $d\in{\Bbb N}$  (see \S 4). 
In particular, one of our main results given in Theorem  4.1,
says that 
\begin{equation}\label{eq:Lgd}
L_{\gamma,d} \le 2 L_{\gamma,d}^{cl}, \qquad 1\le\gamma<3/2,
\quad d\in{\Bbb N},
\end{equation}
whereas for large dimensions it was only known that 
$L_{\gamma,d} \le C \sqrt{d}\, L_{\gamma,d}^{cl}$ 
with some constant 
$C>0$. 

For the important case $\gamma=1$, $d=3$ we have $ L_{1,3}
\le 2L_{1,3}^{cl}< 0.013509$ 
compared with $L_{1,3}< 5.96677 L_{1,3}^{cl}<0.040303$ 
obtained in \cite{L2}
and its improvement
$L_{1,3}< 5.21803 L_{1,3}^{cl}<0.035246$ obtained in \cite{Bla}.

Note also that our estimates on the constant $L_{\gamma,d}$
imply that $L_{1,d}\le2L_{1,d}^{cl}<L_{0,d}^{cl}$ 
as was conjectured in \cite{Rue}. 

In order to obtain our results we give a version of the proof 
obtained in \cite{HLT} for matrix-valued potentials (see \S 3).
Note that E.H.Lieb has informed us that the original
proof obtained in \cite{HLT} 
also works for matrix-valued potentials.
After that in \S 4 we apply the equality
$L_{0,d}=L_{0,d}^{cl}$, for $\gamma\ge3/2$ and $d\in{\Bbb N}$
shown in \cite{LaWe} by using the 
``lifting'' argument with respect to 
the dimension $d$ suggested in \cite{Lap}. 
The same arguments as in \cite{LaWe}
yield the corresponding 
inequalities for Schr\"odinger operators with magnetic fields.
 
In \S 5 we recover the matrix-valued version of the 
Buslaev-Faddeev-Zakharov
trace formulae obtained in \cite{LaWe}
and find some new two sides spectral inequalities
for one-dimensional Schr\"odinger operators with 
operator-valued potentials.

Finally, we are very grateful to L.E.Thomas who was also involved in 
the new proof of Theorem~\ref{Lonehalfone} and has written 
\S~\ref{Larr} as well as reading the text 
of the paper and making many valuable remarks.

\section{Notation and Auxiliary Material}

Let ${\boldsymbol G}$ be a separable Hilbert space with
the norm $\|\cdot\|_{{\boldsymbol G}}$ and the scalar product
$\left<\cdot,\cdot\right>_{{\boldsymbol G}}$ and
let ${\boldsymbol 0}_{{\boldsymbol G}}$ and 
${\boldsymbol 1}_{{\boldsymbol G}}$ be the 
zero and the identity operator on ${\boldsymbol G}$. Denote by
${\mathscr B}({\boldsymbol G})$ the Banach space 
of all bounded operators
on ${\boldsymbol G}$ and by ${\mathscr K}({\boldsymbol G})$ the
(separable) ideal of all compact operators.
Let ${\mathscr S}_1({\boldsymbol G})$ and  
${\mathscr S}_2({\boldsymbol G})$
be the classes of trace  and Hilbert-Schmidt 
operators on ${\boldsymbol G}$ respectively.
For a nonnegative operator $A\in {\mathscr K}({\boldsymbol G})$ 
\begin{equation*}
\lambda_1(A)\geq \lambda_2(A)\geq\ldots\geq 0
\end{equation*}
is the  ordered sequence of its eigenvalues 
(including multiplicities). We use the symbol ``$\tr$''  
to denote traces of operators (matrices) in different Hilbert 
spaces.
 
The Hilbert space ${\boldsymbol H}=L^2({\Bbb R}^d,{\boldsymbol G})$
is the space of all 
measurable functions $u:{\Bbb R}^d\to{\boldsymbol G}$
such that
\begin{equation*}
\|u\|^2_{{\boldsymbol H}}:=\int_{{\Bbb R}^d}\|u\|^2_{{\boldsymbol
G}}\,dx\,
<\infty.
\end{equation*} 
The Sobolev space $H^1({\Bbb R}^d,{\boldsymbol G})$ consists of all
functions $u\in {\boldsymbol H}$ whose norm
\begin{equation*}
\|u\|^2_{H^1({\Bbb R}^d,{\boldsymbol G})}=
\sum_{k=1}^d
\|\partial u/\partial x_k\|^2_{{\boldsymbol H}}+\|u\|^2_{{\boldsymbol
H}}
\end{equation*}
is finite.
Obviously the quadratic form 
\begin{equation*}
h[u,u]=\sum_{k=1}^d \|\partial u/\partial x_k\|^2_{{\boldsymbol H}}
\end{equation*}
is closed in $L^2({\Bbb R}^d,{\boldsymbol G})$ 
on the domain $u\in H^1({\Bbb R}^d,{\boldsymbol G})$.
Let 
\begin{equation*}
V(\cdot):{\Bbb R}^d\to B({\boldsymbol G})
\end{equation*}
be an operator-valued  function satisfying
\begin{equation}\label{V}
\|V(\cdot)\|_{B({\boldsymbol G})} \in L^p({\Bbb R}^d)
\end{equation}
for some finite $p$ with
\begin{alignat*}{5}
p&\geq& 1\quad&\mbox{if}&\quad d&=&1\,,\\
p&>&1\quad&\mbox{if}&\quad d&=&2\,,\\
p&\geq &d/2\quad&\mbox{if}&\quad d&\geq&3\,.
\end{alignat*}
Then the quadratic form 
\begin{equation*}
v[u,u]=\int_{{\Bbb R}^d} \left<V u,u\right>_{{\boldsymbol G}}\,dx
\end{equation*}
is bounded with respect to $h[\cdot,\cdot]$
and thus  the form
\begin{equation}
h[u,u]+v[u,u]
\end{equation}
is closed and semi-bounded from below on $H^1({\Bbb R}^d,{\boldsymbol
G})$.
It generates the self-adjoint operator 
\begin{equation}\label{Q}
Q=-\left(\Delta\otimes{\boldsymbol 1}_{\boldsymbol G}\right) +V(x)
\end{equation}
in $L^2({\Bbb R}^d,{\boldsymbol G})$.
It is not difficult to see, that if   
the operator $V(x)$ belongs to ${\mathscr K}({\boldsymbol G})$
for a.e. $x\in{\Bbb R}^d$ and satisfies the condition~\eqref{V}, 
then the negative spectrum 
\begin{equation*}
-E_1\leq -E_2\leq \cdots \leq -E_n \leq \cdots <0
\end{equation*}
of the operator $Q$ is discrete.

\section{An upper bound for the eigenvalue moment 
in the critical case 
$d=1$ and $\gamma=1/2$.}
\label{section2}

\subsection {A sharp Lieb-Thirring inequality for $d=1$ and
$\gamma=1/2$.}
In this section we give a version of the proof from
\cite{HLT} which will be applied to the Schr\"odinger 
operators with operator-valued potentials. 
The main result of this 
section is the following statement:

\begin{theorem}\label{Lonehalfone}
Let $V(x)$ be a nonpositive operator-valued function,
such that $V(x)\in{\mathscr S}_1({\boldsymbol G})$ 
for a.e. $x\in \rz$ and  $\trG V_-(\cdot)\in L^1({\Bbb R})$. Then
  \begin{equation}\label{g=2}
    \tr
\Big(-\frac{d^2}{dx^2}\!\otimes\! \idG + V\Big)_{-}^{1/2} 
    = \sum_j \sqrt{E_j} \,\leq \frac{1}{2}
    \int_{-\infty}^{\infty} \tr V_{-}\, dx\,.
  \end{equation}
\end{theorem}

\noindent
\textbf{Remark.} The constant 
$L_{1/2,1}=1/2=2 \LTcl_{1/2,1}$ is the best possible. 
Indeed,  $1/2$ is achieved by the operator 
of rank one $V(x) = \delta(x)\left<\cdot, e\right>e$, where 
$e\in {\boldsymbol G}$ and $\delta$ is Dirac's $\delta$-function
(see \cite{HLT}).
\vspace{2mm}

We follow the strategy of \cite{HLT}
quite closely but give a different proof of the
monotonicity lemma.

\subsection{Monotonicity Lemma}
In order to prove the monotonicity lemma we need an 
auxiliary ``majorization'' result. 
Let $A\in{\mathscr K}({\boldsymbol G})$ and let us denote
\begin{equation*}
\|A\|_n=\sum_{j=1}^n \sqrt{\lambda_j(A^*A)}.
\end{equation*}
Then by Ky-Fan's inequality (see for example
\cite[Lemma 4.2]{GK}) the functionals $\|\cdot\|_n$, $n=1,2,\dots$, 
are norms on ${\mathscr K}({\boldsymbol G})$ and thus 
for any unitary operator $U$ in ${\boldsymbol G}$ we have
\begin{equation*}\label{eq:U}
\|U^*AU\|_n=\|A\|_n.
\end{equation*}
\begin{definition}\label{def:majorization}
Let $A$, $B$ be two compact operators on ${\boldsymbol G}$.
We say that $A$ majorizes $B$ or $B \prec A$, iff
  \begin{displaymath}
   \|B\|_n\leq\|A\|_n
   \qquad \text{for all\,\,} n\in\nz\,.
  \end{displaymath}
\end{definition}

\begin{lemma}[Majorization]
\label{majorization}
Let $A$ be a nonnegative compact operator ${\boldsymbol G}$,
$\{U(\omega)\}_{\omega\in\Omega}$
be a family of unitary operators  on ${\boldsymbol G}$, and 
let $g$ be a probability measure on $\Omega$. Then the operator
\begin{displaymath}
  B:= \int_{\Omega} U^*(\omega)AU(\omega)\, g(d\omega) 
  \end{displaymath}
  is majorized by $A$.
\end{lemma}
\begin{proof}
This is a simple consequence of the triangle inequality
\begin{equation*}
\|B\|_n\leq \int_\Omega\|U^*(\omega)AU(\omega)\|_n\,g(d\omega)
=g(\Omega)\|A\|_n=\|A\|_n\,.
\end{equation*}
\end{proof}

\noindent
\textbf{Remark.}
The notion of majorization is well-known in matrix theory
(see \cite{Bh}). 
For finite dimensional Hilbert spaces ${\boldsymbol G}$
even the converse statement of Lemma~\ref{majorization} is
true, cf.\ \cite[Theorem 7.1]{An}: 

\noindent
If $A$ and $B$ are nonnegative  matrices and $\tr A=\tr B$,  
then  the condition $B \prec A$ implies 
that there exist unitary matrices $U_j$
and $t_j>0$, $j=1,\dots,N$, such that
\begin{equation*}
 \sum_{j=1}^N t_j = 1, \qquad B=\sum_{j=1}^N t_j U_j^*AU_j.
\end{equation*}
\vspace{2mm}

Let $W(\cdot): {\Bbb R} \to {\mathscr S}_2({\boldsymbol G})$
be an operator-valued function and let $\|W(\cdot)\|_{{\mathscr S}_2}
\in L^2({\Bbb R})$.
Denote
\begin{equation}
  \label{eq:Loperator}
  \cL_{\varepsilon} :=
 W^* \left[2\varepsilon\Big(-\frac{d^2}{dx^2}+\varepsilon^2\Big)^{-1}
 \otimes\!\idG\right] W\,.
\end{equation}
Obviously, $ \cL_{\varepsilon}$ is a  nonnegative, trace class operator
on 
$L^2({\Bbb R},{\boldsymbol G})$, its trace is independent of 
$\varepsilon$, $0\le \varepsilon<\infty$ and equals 
$\tr \cL_{\varepsilon} = \int \| W(x)\|_{{\mathscr S}_2}^2\, dx$.   
 
\begin{lemma}[Monotonicity]\label{monotonicity}
The operator 
$\cL_\varepsilon$ is majorized by $\cL_{\varepsilon^{\prime}}$
\begin{equation*}
 \cL_\varepsilon \prec \cL_{\varepsilon^{\prime}}
\end{equation*}
 for
all $0\leq \varepsilon^{\prime}\leq \varepsilon$.
\end{lemma}
\begin{proof}
Using the majorization Lemma~\ref{majorization} the proof is basically
reduced  to a  right choice of notation. Let $A$ be the nonnegative
compact operator in $L^2({\Bbb R},{\boldsymbol G})$,  given by the
integral
kernel\footnote{In the scalar case $A$ would just be the rank one
operator
$|W\rangle \langle W|$ (in Dirac
notation).}
$A(x,y):= W^*(x)W(y)$.  
Furthermore let
\begin{equation}
g_{\varepsilon}(dp) = 
\begin{cases}
\varepsilon(\pi(p^2\! +\! \varepsilon^2))^{-1}\,dp
\quad&\text{if}\quad \varepsilon>0\\
\delta(dp)
\quad&\text{if}\quad \varepsilon=0\\
\end{cases}
\end{equation}
be the Cauchy distribution and $\{U(p)\}_{p\in \rz}$ be the group of
unitary multiplication operators $(U(p)\psi)(x) = \e^{-ipx}\psi(x)$ on
$L^2(\rz,\bsG)$. Passing to the Fourier representation of the Green
function in~\eqref{eq:Loperator} we obtain
\begin{equation}\label{eq:representation}
  \cL_{\varepsilon} = 
  \int_{-\infty}^{\infty} U^*(p) A U(p)
  \, g_\varepsilon(dp)\,.
\end{equation}
Of course, $\cL_0 = A$. In particular, 
Lemma~\ref{majorization} and~\eqref{eq:representation}
immediately imply $\cL_{\varepsilon} \prec\cL_{0}$. 
The Cauchy distribution is a convolution semigroup, i.e.
$g_\varepsilon =
g_\varepsilon{^{\prime}}*g_{\varepsilon-\varepsilon^{\prime}}$. 
If we insert this into 
\eqref{eq:representation} and change variables using the group
property of the unitary operators $U(p)$, then  Lemma~\ref{majorization}
yields
\begin{displaymath}
  \cL_{\varepsilon} = 
  \int U^*(p)\cL_{\varepsilon^{\prime}} U(p)
  \, g_{\varepsilon- \varepsilon^{\prime}}(p) dp
  \prec \cL_{\varepsilon^{\prime}}.
\end{displaymath}
This completes the proof.
\end{proof}

\subsection{Proof of Theorem~\ref{Lonehalfone}}
Let $W(x) = \sqrt{V_{-}(x)}$, so $W^*=W$. Then from the assumptions
made in Theorem~\ref{Lonehalfone}, we find that $W(x)$ is a 
family of nonnegative Hilbert-Schmidt operators such that 
$\|W(\cdot)\|_{{\mathscr S}_2}\in L^2({\Bbb R})$. Let 
\begin{equation}
  \label{eq:BSoperator}
  \cK_{E} := \frac{1}{2\sqrt{E}} \cL_{\sqrt{E}} =
  W \left[\Big(-\frac{d^2}{dx^2}+E\Big)^{-1}\otimes\!\idG\right] W\,,
\end{equation}
where $ \cL_{\varepsilon}$ is defined in (\ref{eq:Loperator}).
According to the
Birman-Schwinger principle \cite{Bir,Schw}
we have
\begin{equation*}
1=\lambda_{j}(\cK_{E_{j}})
\end{equation*} 
for all negative eigenvalues $\{-E_j\}_j$ of the Schr\"odinger operator
(\ref{Q}).
Multiplying this equality
by $2\sqrt{E_{j}}$ and summing over $j$ we obtain
\begin{equation}
  \label{eq:basic}
  2\sum \sqrt{E_j} = \sum \lambda_j(\cL_{\sqrt{E_j}}).
\end{equation}
In contrast to $\cK_{E}$ the operator 
$\cL_{\sqrt E}$ is well-behaved for small energies.
We now use the same 
\emph{monotonicity argument} as in 
\cite{HLT} to dispose of the energy dependence of the operator in 
\eqref{eq:basic}. Namely, for any $n\in \nz$, 
Lemma~\ref{monotonicity} implies that
the \emph{partial traces}  $\sum_{j\leq n} \lambda_j(\cL_{\varepsilon})$ 
are \emph{monotone decreasing} in $\varepsilon$. 
Given this monotonicity, a simple induction 
argument yields
\begin{displaymath}
  \sum_{j\leq n} \lambda_j(\cL_{\sqrt{E_j}})
  \leq 
  \sum_{j\leq n} \lambda_j(\cL_{\sqrt{E_n}})
  \quad \text{for all } n\in \nz.
\end{displaymath}
Hence, by~\eqref{eq:basic}
we also have the
bound  
\begin{displaymath}
  2\sum \sqrt{E_j} \leq 
  \sum \lambda_j(\cL_{0}) = \tr \cL_{0} = \int_{-\infty}^{\infty}
\tr W^2(x)\, dx = 
  \int_{-\infty}^{\infty} \trG V_{-}(x)\, dx.
\end{displaymath}
The proof is complete.

\subsection{Some generalizations of
Theorem~\ref{Lonehalfone}}\label{Larr}
The above strategy can be adapted to obtain upper bounds 
on eigenvalue moments for operators of the form 
$H= |-i\nabla|^{\beta}+V$ acting in $L^{2}(R^{d})$, 
$\beta>d$.  Suppose that $\Phi$ is an infinitely divisible 
symmetric probability density, e.g.\ a compound Poisson, 
of the form 
\begin{equation*} 
  \Phi(p)=\frac{1}{(2\pi)^{d}}\int e^{\int(\cos(x\cdot\xi)-1)dm(\xi)}
e^{ip\cdot x}dx 
\end{equation*}
with $m$ a non-negative measure and such that $\Phi$ 
satisfies a point-wise inequality
\begin{equation}\label{t2}
 \left(|p|^{\beta} +1\right)^{-1}\leq c_{0}\Phi(p)
\end{equation}
for some constant $c_{0}$.  Then by scaling,   
\begin{equation*}
E^{\frac{\beta-d}{\beta}}\left(|p|^{\beta}+E\right)^{-1}\leq 
c_{0}E^{-d/\beta}\Phi(p/E^{\beta})\equiv \Psi_E(p).
\end{equation*}
Moreover, 
\begin{equation*}
   \Psi_E(p)\equiv \Theta_{E,E^{\prime}}* \Psi_{E^{\prime}}(p)
\end{equation*}
where $\Theta_{E,E^{\prime}}$ is a non-negative probability 
density with Fourier transform given by 
\begin{equation*}
  \hat{\Theta}_{E,E^{\prime}}(x)=
\exp{\left\{\int(\cos(x\cdot\xi)-1)[dm(\xi/E^{1/\beta})
-dm(\xi/E^{\prime\,1/\beta})]\right\}}
\end{equation*}
{\it provided} that $[dm(\xi/E^{1/\beta})-dm
(\xi/E^{\prime\,1/\beta})]$ 
is non-negative for $E^{\prime}\leq  E$.  

Assuming that $dm$ satisfies this condition, 
we have by the majorization argument that
\begin{eqnarray}\label{t6}
\sum_{j\leq n}E_{j}^{(\beta-d)/\beta}(H)\!\!&\leq& \sum_{j\leq
n}\lambda_j(V_{-}^{1/2}\frac{E_{j}^{(\beta-d)/\beta}}
{|-i\nabla|^{\beta}+E_j}V_{-}^{1/2})\nonumber\\
&\leq& \sum_{j\leq n}\lambda_j(V_{-}^{1/2}\Psi_{E_j}
(-i\nabla)V_{-}^{1/2})\\&\leq& {\rm tr}\,(V_{-}^{1/2}
\Psi_{E_n}(-i\nabla)V_{-}^{1/2}) 
= \frac{c_{0}}{(2\pi)^d}\int V_{-}(x)dx.\nonumber
\end{eqnarray}

The problem of finding such an optimal $\Phi$ and $c_{0}$ seems
non-trivial in general. But in $d$ dimensions, with the choice
$dm(\xi)=cd\xi/|\xi|^{d+\alpha}$, with $d+\alpha\leq \beta$,
$0<\alpha<2$, $(\cos(x\cdot\xi)-1)$ is integrable with respect to $m$,
and $\int(\cos(x\cdot\xi)-1)dm(\xi)=-c_{1} |x|^{\alpha}$ for some
$c_{1}>0$.  Consequently, $\Phi(p)\sim |p|^{-(\alpha+d)}$,
$p\rightarrow\infty$, and $c_{0}\Phi$ will majorize
$(|p|^{\beta}+1)^{-1}$ for sufficiently large $c_{0}$.  An eigenvalue
moment bound (\ref{t6}) follows.  For the $d=1,\, \beta=2$ 
Cauchy density case
above, the optimal choice is $dm(\xi)= d\xi/(\pi\xi^2)$ and $c_{0}=
\pi$; (\ref{t2}) is an equality.

\subsection{A priori estimate for moments $\mathbf{\gamma\geq 1/2}$}
Following Aizenman and Lieb \cite{AL} we can ``lift'' the bound of 
Theorem~\ref{Lonehalfone} to moments $\gamma \geq 1/2$. 

\begin{corollary}\label{Lonehalfone1}
  Assume that $V(x)$ is a nonpositive operator-valued function 
for a.e. $x\in \rz$ and that 
$\trG V_-(\cdot)\in L^{\gamma+\frac{1}{2}}({\Bbb R})$ for some
$\gamma\geq 1/2$. Then
  \begin{equation}\label{g>2}
    \tr
\Big(-\frac{d^2}{dx^2}\!\otimes\! \idG + V\Big)_{-}^{\gamma} 
    = \sum_j E_j^\gamma \,\leq 2\LTcl_{\gamma,1}
    \int_{-\infty}^{\infty} \tr V_{-}^{\gamma+\frac{1}{2}}\, dx\,.
  \end{equation}
\end{corollary}

\begin{proof}
Note that Theorem~\ref{Lonehalfone} is 
equivalent to 
\begin{equation*}
\tr \Big(-\frac{d^2}{dx^2}\otimes\idG + V\Big)_{-}^{1/2}
  \leq 2 \iint_{{\rz\times\rz}} \trG(p^2 -V_{-}(x))_{-}^{1/2}\,
\frac{dpdx}{2\pi}\,.
\end{equation*}
Scaling gives the simple identity for all $s\in \rz$ 
\begin{equation*}
s_{-}^\gamma=C_\gamma\int_0^\infty t^{\gamma - \frac{3}{2}} 
(s+t)_{-}^{1/2}\, dt\,,\qquad 
C_\gamma^{-1}=B\Big(\gamma-\frac{1}{2},\frac{3}{2}\Big)\,,
\end{equation*}
where $B$ is the Beta function.
Let  $\mu_j(x)$ the eigenvalues of $V_{-}(x)$.
Then 
\begin{align*}
  \tr \Big(-\frac{d^2}{dx^2}\!\otimes\!\idG &+V\Big)_{-}^\gamma 
  = C_\gamma
  \int_0^\infty dt\,t^{\gamma - \frac{3}{2}} 
  \tr \Big(-\frac{d^2}{dx^2}\!\otimes\!\idG +V+t\Big)_{-}^{1/2}
  \\
  &\leq C_\gamma 
    \int_0^\infty dt\,t^{\gamma-\frac{3}{2}} \,
      2 \iint \trG(p^2 -V_{-}+t)_{-}^{1/2}\,\frac{dpdx}{2\pi} 
  \\
  &= 
  2 \sum_{j=1}^{\infty} \iint 
  \left[
      C_\gamma
      \int_0^\infty dt\,t^{\gamma-\frac{3}{2}} (p^2
-\mu_{j}+t)_{-}^{1/2}
    \right]\,\frac{dpdx}{2\pi}
  \\
  &=
  2 \iint \trG (p^2 -V_{-})_{-}^{\gamma}\, \frac{dpdx}{2\pi}
  =
  2\, \LTcl_{\gamma,1} \int \trG V_{-}^{\gamma + 1/2} \, dx\,.
\end{align*}
\end{proof}

\section{New estimates on the constants $L_{\gamma,d}$
for $1/2\le\gamma<3/2$, $d\in{\Bbb N}$}

\subsection{The Main result.} We consider now the Schr\"odinger
operator~\eqref{Q}
in $L^2({\Bbb R}^d,{\boldsymbol G})$
for an arbitrary $d\in\nz$.
Assume that $V$ is a nonpositive operator-valued function satisfying 
the condition   
\begin{equation}\label{trest}
\tr V(\cdot) \in L^{\gamma+\frac{d}{2}}({\Bbb R}^d)
\end{equation}
for some appropriate $\gamma$. We shall discuss bounds on the optimal
constants in the Lieb-Thirring inequalities
\begin{equation}\label{basin}
\tr (-\Delta\otimes{\boldsymbol 1} + V)_-^\gamma
\leq L_{\gamma,d} \int_{{\Bbb R}^d} \tr V_-^{\frac{d}{2}+\gamma}\,dx\,.
\end{equation}
In \cite{LaWe} it has been shown that
\begin{equation}\label{dgeq3/2}
L_{\gamma,d}=L_{\gamma,d}^{\mbox{\footnotesize cl}}
\qquad\mbox{for all}\quad \gamma\geq 3/2,\quad d\in{\Bbb N}\,.
\end{equation}
The main result of the paper concerns  $1/2\leq \gamma <3/2$.

\begin{theorem}\label{t3}
Let $V$ be a nonpositive operator-valued function and let the 
condition~\eqref{trest} be satisfied. Then 
the following estimates on the sharp constants $L_{\gamma,d}$ 
hold
\begin{alignat}{7}
\label{dinN1<g<3/2}
L_{\gamma,d}&\leq& 2 L_{\gamma,d}^{\mbox{\upshape\footnotesize cl}}
\qquad&\mbox{for all}&\quad 1\leq&\gamma&<3/2\,,\quad d&\in&{\Bbb
N}\,,\\
\label{d=11/2<g<3/2}
L_{\gamma,d}&\leq& 2 L_{\gamma,d}^{\mbox{\upshape\footnotesize cl}}
\qquad&\mbox{for all}&\quad 1/2\leq&\gamma&<3/2\,,\quad d&=&1\,,\\
\label{d>11/2<g<1}
L_{\gamma,d}&\leq& 4 L_{\gamma,d}^{\mbox{\upshape\footnotesize cl}}
\qquad&\mbox{for all}&\quad 1/2\leq&\gamma&<1\ \ \,,\quad d&\geq&2\,.
\end{alignat}
\end{theorem}

\noindent 
{\bfseries Remark.}
For the special case $\gamma=1$ we find that 
\begin{equation*}
L_{1,d}^{\mbox{\footnotesize cl}}\leq L_{1,d}\leq 2
L_{1,d}^{\mbox{\footnotesize cl}}
\quad\mbox{for all}\quad d\in{\Bbb N}\,.
\end{equation*}
Even in the scalar case ${\boldsymbol G}={\Bbb C}$ this is a substantial
improvement of the previously 
known numerical estimates on these constants in high dimensions
obtained  
in \cite{Bla} and \cite{L2}.
 
\vspace{2mm}

\noindent
\textbf{Remark.}
In fact, our proof of Theorem~\ref{t3} yields 
\begin{equation*}
L_{\gamma,d}\leq 
\frac{L_{\gamma,1}}{\LTcl_{\gamma,1}}\LTcl_{\gamma,d},\qquad d\in{\Bbb
N}\,,
\quad 1\leq\gamma<3/2\,.
\end{equation*}
According to Corollary~\ref{Lonehalfone1} we know that 
$L_{1,1}\leq 2 \LTcl_{1,1}$. In the scalar case Lieb and Thirring
conjectured that 
\begin{equation*}
\frac{L_{\gamma,1}}{\LTcl_{\gamma,1}}=2\left(\frac{\gamma-1/2}
{\gamma+1/2}\right)^{\gamma-1/2}\,,
\qquad
1/2\leq \gamma <3/2\,.
\end{equation*}
In particular,
if this were true in the matrix case for $\gamma=1$,
our approach would imply 
$\LTcl_{1,1}\leq L_{1,d}< 1.16\, \LTcl_{1,d}$.

\begin{proof}[Proof of Theorem~\ref{t3}]
We apply an induction argument similar to the one used in
\cite{LaWe}. For $d=1$ and  $1/2\leq \gamma<3/2$
the bound~\eqref{d=11/2<g<3/2} is identical to~\eqref{g>2}.

Consider the operator~\eqref{Q}
in the (external) dimension $d$. 
We rewrite the quadratic form $h[u,u]+v[u,u]$ 
for $u\in H^1({\Bbb R}^d,{\boldsymbol G})$ as 
\begin{align*}
h[u,u]+v[u,u]&=\int_{-\infty}^{+\infty} h(x_d)[u,u]\,dx_d
+\int_{-\infty}^{+\infty} w(x_d)[u,u]\,dx_d\,,\\
h(x_d)[u,u]&=\int_{{\Bbb R}^{d-1}}
\left\|\frac{\partial u}{\partial x_d}\right\|^2_{\boldsymbol
G}dx_1\cdots x_{d-1}\,,\\
w(x_d)[u,u]&=\int_{{\Bbb R}^{d-1}}
\left[
\sum_{j=1}^{d-1}\left\|\frac{\partial u}{\partial
x_j}\right\|^2_{\boldsymbol G}
+
\left<V(x)u,u\right>_{\boldsymbol G}
\right]
dx_1\cdots x_{d-1}\,.
\end{align*}
The form $w(x_d)$ is closed on $H^1({\Bbb R}^{d-1},{\boldsymbol G})$ for
a.e.
$x_d\in{\Bbb R}$ and 
it induces the self-adjoint operator
\begin{equation*}
W(x_d)=-\sum_{k=1}^{d-1}\frac{\partial^2}{\partial x_k^2}
\otimes \idG+V(x_1,\dots,x_{d-1};x_d) 
\end{equation*}
on $L^2({\Bbb R}^{d-1},{\boldsymbol G})$. 
For a fixed $x_d\in{\Bbb R}$ this is a Schr\"odinger operator
in $d-1$ dimensions. Its
negative spectrum 
is discrete, hence $W_-(x_d)$ is compact
on $L^2({\Bbb R}^{d-1},{\boldsymbol G})$.

Assume that we have 
\eqref{dinN1<g<3/2}--\eqref{d=11/2<g<3/2} 
for  the dimension $d-1$ and all $\gamma$ from the interval
$1/2\leq\gamma<3/2$. 
Then
$\tr W_-^{\gamma+\frac{1}{2}}(x_d)$ satisfies the bound
\begin{equation}\label{eq:trwww}
\tr W^{\gamma+\frac{1}{2}}_-(x_d)
\leq L_{\gamma+\frac{1}{2},d-1}\int_{{\Bbb R}^{d-1}}
\tr V_-^{\gamma+\frac{d}{2}}(x_1,\dots,x_{d-1};x_d)
\,dx_1\cdots dx_{d-1}
\end{equation}
for a.e. $x_d\in{\Bbb R}$. Here 
\begin{align}
\label{gg>1}
L_{\gamma+\frac{1}{2},d-1}
=L_{\gamma+\frac{1}{2},d-1}^{\mbox{\footnotesize cl}}
\qquad&\mbox{for}\qquad
\gamma\geq 1\,,\\
\label{gg<1}
L_{\gamma+\frac{1}{2},d-1}
\leq 2L_{\gamma+\frac{1}{2},d-1}^{\mbox{\footnotesize cl}}
\qquad&\mbox{for}\qquad
1/2\leq \gamma< 1\,.
\end{align}
Indeed, \eqref{gg>1} follows from~\eqref{dgeq3/2} and~\eqref{gg<1} 
follows from~\eqref{dinN1<g<3/2}--\eqref{d=11/2<g<3/2}
in dimension $d-1$.

Let $w_-(x_d)[\cdot,\cdot]$ be the quadratic form corresponding to
the operator $W_-(x_d)$ on 
${\boldsymbol H}=L^2({\Bbb R}^{d-1},{\boldsymbol G})$.
We have $w(x_d)[u,u]\geq -w_-(x_d)[u,u]$ and
\begin{equation}\label{eq:rhs}
h[u,u]+v[u,u]\geq \int_{-\infty}^{+\infty}
\left[
\left\|\frac{\partial u}{\partial x_d}\right\|^2_{\boldsymbol H}
-\left<W_-(x_d)u,u\right>_{\boldsymbol H}
\right]\,dx_d
\end{equation}
for all $u\in H^1({\Bbb R}^{d},{\boldsymbol G})$. 
According to section 2.2 the form on the r.h.s.\ of~\eqref{eq:rhs}
can be closed to
$H^1({\Bbb R},{\boldsymbol H})$ and induces the self-adjoint operator
\begin{equation*}
-\frac{d^2}{dx_d^2}\otimes{\boldsymbol 1}_{\boldsymbol H} -
W_-(x_d)
\end{equation*}
on $L^2({\Bbb R},{\boldsymbol H})$.
Then~\eqref{eq:rhs} implies
\begin{equation}\label{eq:llaass}
  \tr(-\Delta\otimes{\boldsymbol
    1}_{\boldsymbol G}+V)_-^\gamma \leq \tr\left(
    -\frac{d^2}{dx_d^2}\otimes{\boldsymbol 1}_{\boldsymbol H} -
W_-(x_d)\right)_-^\gamma \,.
\end{equation}
The assumption $V\in L^{\gamma+\frac{d}{2}}({\Bbb R}^{d})$
implies that $\tr W_-^{\gamma+\frac{1}{2}}$ is an
integrable function and we
can apply Corollary~\ref{Lonehalfone1} to the r.h.s.\ of
\eqref{eq:llaass}.
In view of~\eqref{eq:trwww} we find
\begin{align*}
\tr\left(
    -\frac{d^2}{dx_d^2}\otimes{\boldsymbol 1}_{\boldsymbol H} -
    W_-(x_d)\right)_-^\gamma
  &\leq L_{\gamma,1}\int_{-\infty}^{+\infty}\tr
   W_-^{\gamma+\frac{1}{2}}(x_d)\,dx_d\\
  &\leq L_{\gamma,1}L_{\gamma+\frac{1}{2},d-1} \int_{{\Bbb
      R}^d} \tr V_-^{\gamma+\frac{d}{2}}\,dx
\end{align*}
for $\gamma\geq 1/2$. The bounds~\eqref{d=11/2<g<3/2},~\eqref{gg>1}
or~\eqref{gg<1} and the calculation
\begin{align*}
  L^{\mbox{\footnotesize\upshape
cl}}_{\gamma,1}L^{\mbox{\footnotesize\upshape
cl}}_{\gamma+\frac{1}{2},d-1}
  &= \frac{\Gamma(\gamma+1)}
{2\pi^{\frac{1}{2}}\Gamma(\gamma+\frac{1}{2}+1)}
  \cdot \frac{\Gamma(\gamma+\frac{1}{2}+1)}
  {2^{d-1}\pi^{\frac{d-1}{2}}
  \Gamma(\gamma+ \frac{1}{2}+\frac{d-1}{2}+1)}\\
  &=
\frac{\Gamma(\gamma+1)}{2^d\pi^{\frac{d}{2}}\Gamma(\gamma+\frac{d}{2}+1)}
  =L^{\mbox{\footnotesize\upshape cl}}_{\gamma,d}
\end{align*}
complete the proof.  
\end{proof}

\subsection{Estimates for magnetic Schr\"odinger operators}
Following a remark by B. Helffer \cite{Hel} 
and using the arguments from \cite{LaWe} we can 
extend Theorem~\ref{t3} to Schr\"odinger operators with magnetic
fields.  
Let $Q({\boldsymbol a})$ be a self-adjoint operator in 
$L^2({\Bbb R}^d,{\boldsymbol G})$
\begin{equation}\label{Qa}
Q({\boldsymbol a})=(i\nabla+{\boldsymbol a}(x))^2
\otimes{\boldsymbol 1}_{\boldsymbol G}+V(x),
\end{equation}
where  
\begin{equation*}
{\boldsymbol a}(x)=(a_1(x), \cdots, a_d(x))^t\,,\qquad d\geq 2\,,
\end{equation*}
is a magnetic vector potential with real-valued entries
$a_k\in L^2_{\footnotesize loc}({\Bbb R}^d)$.

We consider the inequality
\begin{equation}\label{mglt}
\tr (Q({\boldsymbol a}))_-^\gamma \leq \tilde{L}_{\gamma,d}
\int_{{\Bbb R}^d} V_-^{\frac{d}{2}+\gamma}\,dx\,,
\end{equation}
where the nonpositive operator function $V(\cdot)$
satisfies~\eqref{trest}.
In \cite{LaWe} it has been shown, that
\begin{equation}\label{mgcl}
\tilde{L}_{\gamma,d}=L_{\gamma,d}^{\mbox{\footnotesize cl}}
\qquad\mbox{for all}\qquad
\gamma\geq 3/2\,,\quad d\in{\Bbb N}\,.
\end{equation}
In general, the sharp constant $\tilde{L}_{\gamma,d}$ in~\eqref{mgcl} 
might differ from the sharp constant
$L_{\gamma,d}$ in~\eqref{basin}
\begin{equation*}
L_{\gamma,d}^{\mbox{\footnotesize cl}}
\leq
L_{\gamma,d}
\leq
\tilde{L}_{\gamma,d}\,.
\end{equation*}
By combining the arguments from \cite{LaWe} and 
those used in the prove of Theorem~\ref{t3}
we immediately obtain the following result:

\begin{theorem}\label{t4}
The following estimates on the sharp constants $\tilde{L}_{\gamma,d}$ 
in~\eqref{mglt} hold
\begin{alignat}{7}
\label{mgdinN1<g<3/2}
\tilde{L}_{\gamma,d}&\leq& 2 L_{\gamma,d}^{\mbox{\upshape\footnotesize
cl}}
\qquad&\mbox{for all}&\quad 1\leq&\gamma&<3/2\,,\quad d&\geq&2\,,\\
\label{mg>11/2<g<1}
\tilde{L}_{\gamma,d}&\leq& 4 L_{\gamma,d}^{\mbox{\upshape\footnotesize
cl}}
\qquad&\mbox{for all}&\quad 1/2\leq&\gamma&<1\ \ \,,\quad d&\geq&2\,.
\end{alignat}
\end{theorem}
 
\section{Trace formulae and estimates from below for $d=1$.}

\subsection{Matrix-valued potentials}
Let ${\boldsymbol G}={\Bbb C}^n$ be a finite dimensional Hilbert space.
We consider the system of ordinary differential equations
\begin{equation}\label{y}
-\left(\frac{d^2}{dx^2}\otimes {\boldsymbol
1}\right)y(x)+V(x)y(x)=k^2y(x)\,,
\qquad x\in{\Bbb R}\,,
\end{equation}
where $V$ is a compactly supported, smooth 
(not necessary sign definite) Hermitian matrix-valued
function. Define  
\begin{equation*}
x_{\min}:=\min \mbox{supp\ } V
\quad\mbox{and}\quad 
x_{\max}:=\max\mbox{supp\ } V\,.
\end{equation*}
Then for any $k\in{\Bbb C}\backslash\{0\}$  there exist 
unique $n\times n$ matrix-solutions $F(x,k)$ and $G(x,k)$ of the
equations
\begin{align}
\label{Feq}
-F^{\prime\prime}_{xx}(x,k)+VF(x,k)&=k^2F(x,k)\,,\\
\label{Geq}
-G^{\prime\prime}_{xx}(x,k)+VG(x,k)&=k^2G(x,k)\,,
\end{align}
satisfying
\begin{alignat*}{4}
F(x,k)&=e^{ikx} {\boldsymbol 1}_{\bsG} 
\quad&\mbox{as}\quad x&\geq x_{\max}\,,\\
G(x,k)&=e^{-ikx}{\boldsymbol 1}_{\bsG} 
\quad&\mbox{as}\quad x&\leq x_{\min}\,.
\end{alignat*}
If $k\in{\Bbb C}\setminus\{0\}$, then the pairs 
of matrices $F(x,k)$, $F(x,-k)$ and $G(x,k)$, $G(x,-k)$ form
full systems of independent solutions of \eqref{y}.
Hence the matrix
$F(x,k)$ can be expressed as a linear combination of $G(x,k)$ and
$G(x,-k)$
\begin{equation}\label{1}
F(x,k)=G(x,k)B(k)+G(x,-k)A(k)\,.
\end{equation}
The matrix functions $A(k)$ and $B(k)$ are uniquely defined by
\eqref{1}.

\subsection{Trace formulae}
In \cite{LaWe} the Buslaev-Faddeev-Zakharov 
trace formulae were generalized for the matrix-valued potentials 
satisfying the conditions from the previous subsection. 
We recall here
the first three trace identities given by the
equations (1.60)-(1.62) from \cite{LaWe} 
\begin{align}\label{final1}
  \frac{1}{4}\int_{-\infty}^{+\infty}\tr V\,dx &=
  I_0
  -\sum_{l=1}^N E^{1/2}_l\,,\\
\label{final2}
\frac{3}{16} \int_{-\infty}^{+\infty}\tr V^2\,dx &=
  3 I_2 +
\sum_{l=1}^N E^{3/2}_l\,,\\
\label{final3}
\frac{5}{32} \int_{-\infty}^{+\infty}\tr V^3\,dx
+\frac{5}{64} \int_{-\infty}^{+\infty}\tr
\left(\frac{dV}{dx}\right)^2\,dx 
&=
  5 I_4 -
\sum_{l=1}^N E^{5/2}_l\,,
\end{align}
where 
\begin{equation*}
I_j=(2\pi)^{-1}
\int_{-\infty}^{+\infty}k^j \ln |\det A(k)|\,dk\,\qquad j=0,2,4\,.
\end{equation*}
Note that for real $k$'s we have (cf. (1.11) in \cite{LaWe})
\begin{equation*}
A(k)A^*(k)={\boldsymbol 1}_{\bsG}+B(-k)B^*(-k)
\end{equation*}
Thus we obtain
$|\det A(k)|\geq 1$ for all $k\in{\Bbb R}$ 
and 
\begin{equation}\label{lj}
I_{j}\geq 0\,\qquad j=0,2,4\,.
\end{equation}

\noindent
\textbf{Remark.}
Notice that
\begin{equation}\label{constants}
L_{1/2,1}^{\mbox{\footnotesize cl}}=1/4\,,
\qquad
L_{3/2,1}^{\mbox{\footnotesize cl}}=3/16\,,
\qquad
L_{5/2,1}^{\mbox{\footnotesize cl}}=5/32\,.
\end{equation}

\subsection{$\gamma=1/2$} 
The identity~\eqref{final1}  immediately leads to a bound from below
on the sum of the square roots of the operator~\eqref{y}.
Indeed,  \eqref{lj} implies 
\begin{equation}\label{below}
L_{1/2,1}^{\mbox{\footnotesize cl}}
\int (\tr V_- - \tr V_+)\,dx \leq 
\sum_{l} E^{1/2}_l.
\end{equation}
For the scalar case this estimate has been pointed out
in \cite{GGM}, see also \cite{W1}.
By continuity this bound extends to all matrix functions
$V$, for which
\begin{equation}\label{tr+-}
\tr V_+(\cdot) \in L^1({\Bbb R}) \qquad \mbox{and}\qquad  \tr
V_-(\cdot)\in L^1({\Bbb R})\,.
\end{equation}
Using a standard density argument and~\eqref{g=2}
we conclude, that~\eqref{below} holds also for general separable
Hilbert spaces ${\boldsymbol G}$. This implies
\begin{corollary}
Let $V(x)\in S_1(\bsG)$ and $\tr V_\pm(\cdot)\in L_1({\Bbb R})$.
Then for the 1/2 moments of the negative eigenvalues
of the operator~\eqref{Q} we have the following two side inequalities
\begin{equation*}
\LTcl_{1/2,1}
\int
(\tr V_--\tr V_+)\,dx 
\leq \sum_{l} E_l^{1/2}
\leq 2\LTcl_{1/2,1}
\int
\tr V_-\,dx.
\end{equation*}
\end{corollary}

\subsection{${\boldsymbol \gamma=3/2}$}
Let us return to the case
${\boldsymbol G}={\Bbb C}^n$ and let $V$ be a smooth, 
compactly supported matrix-valued function. 
The upper bound~\eqref{g=2} and
the identity~\eqref{final1} imply
\begin{align}\label{bfa}
I_1&=\sum_{l=1}^N E_l^{1/2} + L_{1/2,1}^{\mbox{\footnotesize
    cl}}
\int (\tr V_+-\tr V_-)\,dx\\
\notag
&\leq 
 L_{1/2,1}^{\mbox{\footnotesize cl}}
\int (\tr V_++\tr V_-)\,dx\,.
\end{align}
Moreover,
from \eqref{dgeq3/2} with $d=1$ and $\gamma=5/2$, 
\eqref{final3} and~\eqref{constants}
it follows that 
\begin{align}\notag
5 I_4 
&= L_{5/2,1}^{\mbox{\footnotesize cl}}
\int (\tr V_+^3-\tr V_-^3)\,dx +
\frac{1}{2}L_{5/2,1}^{\mbox{\footnotesize cl}}
\int\tr \left(\frac{dV}{dx}\right)^2dx
+\sum_l E_l^{5/2}
\\
\label{fba}
&\leq L_{5/2,1}^{\mbox{\footnotesize cl}}
\int \tr V_+^3\,dx +
\frac{1}{2}L_{5/2,1}^{\mbox{\footnotesize cl}}
\int\tr \left(\frac{dV}{dx}\right)^2\,dx\,. 
\end{align}
Note that in the scalar case the inequalities~\eqref{bfa}
and~\eqref{fba} with somewhat worse constant were found in 
\cite{W1} and \cite{LT} respectively.
These estimates together with 
H\"older's inequality  give 
\begin{equation}\label{i2}
I_2\leq  I_0^{1/2} I_4^{1/2} = \frac{1}{16}
\left[\int (\tr V_++\tr V_-)\,dx\right]^{\frac{1}{2}}
\left[2\int\tr V_+^3\,dx+
\int\tr \left(\frac{dV}{dx}\right)^2\,dx\right]^{\frac{1}{2}}. 
\end{equation}
Inserting~\eqref{i2}  into~\eqref{final2} and considering 
the special case $V_+=0$,  we find
\begin{equation}\label{3/2bel}
\frac{3}{16}
\int
\tr V_-^2\,dx-
\sum_{l} E_l^{3/2}
\leq \frac{3}{16}
\left[\int \tr V_-\,dx\right]^{\frac{1}{2}}
\left[\int\tr \left(\frac{dV}{dx}\right)^2\,dx\right]^{\frac{1}{2}}.
\end{equation}
Standard density and continuity arguments allow us to
extend~\eqref{3/2bel}
to general separable Hilbert spaces ${\boldsymbol G}$ and 
arbitrary nonpositive
operator-valued potentials $V$, for which all integrals
in~\eqref{3/2bel}
are finite. 

\subsection{A remainder term}
Let us discuss further the inequality~\eqref{3/2bel}.
First note, that in view of~\eqref{dgeq3/2} for $d=1$ and $\gamma=3/2$,
the l.h.s.\ of~\eqref{3/2bel} is nonnegative. 
Therefore the inequalities~\eqref{3/2bel} can be interpreted 
as an estimate on the difference between 
the sum $\sum_l E_l^{3/2}$ and
the classical phase space integral
\begin{equation*}
L_{3/2}^{{\mbox{\footnotesize cl}}} \int \tr V_-^{2} dx
=
\iint \tr (p^2+V(x))_-^{3/2} \frac{dpdx}{2\pi}\,.
\end{equation*}

By replacing $V$ by $\alpha V$ 
we obtain the following result:

\begin{theorem} \label{Weylth}
Assume that $V$ is a nonpositive operator-valued function such that
$\tr V_-\in L^1({\Bbb R})\cap L^2({\Bbb R})$
and $\tr (dV/dx)^2 \in L^1({\Bbb R})$. Then
\begin{equation*}
\tr \Big(-\frac{d^2}{dx^2}\!\otimes\!\idG+\alpha V\Big)_-^{3/2}
=\alpha^2 \LTcl_{3/2,1}
\int \tr V_-^2\,dx - R(\alpha)
\end{equation*}
for all $\alpha>0$, where
\begin{equation*}
0\leq R(\alpha) \leq \frac{3\alpha^{3/2}}{16}
\left[\int \tr V_-\,dx\right]^{\frac{1}{2}}
\left[\int\tr \left(\frac{dV}{dx}\right)^2\,dx\right]^{\frac{1}{2}}\,.
\end{equation*}
\end{theorem}

\noindent
\textbf{Remark.}
For large values of the coupling constant $\alpha$, 
Theorem~\ref{Weylth} 
gives us the correct order $O(\alpha^{3/2})$ of the remainder
term in the Weyl asymptotic formula for $3/2$-moments 
of the negative eigenvalues.
 
\subsection{Acknowledgements}
The second and the fourth authors wish to express their 
gratitude to B.Helffer for his valuable comments on
magnetic Schr\"odinger operators.
D.Hundertmark thanks the Mathematical Department of 
the Royal Institute of Technology in Stockholm  
for its warm hospitality and the Deutsche Forschungsgemeinschaft for
financial support under grant Hu 773/1-1. 
A.Laptev has been supported by the Swedish Natural 
Sciences Research Council, Grant M-AA/MA 09364-320, 
T.Weidl has been supported by the Swedish Natural Science
Council dnr 11017-303.

\noindent
Partial financial support from
the European Union through the TMR network FMRX-CT 96-0001 is
gratefully acknowledged.

{\small
\begin{multicols}{2}
  \raggedright
  Departments of Physics, Jadwin Hall$^1$\\
  Princeton University\\
  Princeton, New Jersey 08544, U.S.A\\
  Royal Institute of Technology$^2$\\
  Department of Mathematics\\
  S-10044 Stockholm, Sweden\\
\end{multicols}}

{\small
\begin{multicols}{2}
\noindent
 Universit\"at Regensburg$^3$\\
  Naturwissenschaftliche Fakult\"at I\\
  D-93040 Regensburg, Germany
\end{multicols}}

\end{document}